\documentclass[11pt]{article}
\begin{document}
\date{\today}
\pagestyle{plain}
\title{Thomas Precession by Uniform  Acceleration}
\author{Miroslav Pardy \\ [2mm]
Department of Physical Electronics \\
Masaryk University \\
Kotl\'{a}\v{r}sk\'{a} 2, 611 37 Brno, Czech Republic\\
e-mail:pamir@physics.muni.cz}
\date{\today}
\maketitle

\vspace{20mm}

\begin{abstract}
We determine nonlinear transformations between coordinate
systems which are mutually in a constant symmetrical accelerated motion.
The maximal acceleration limit follows from the kinematical origin 
and it is an analogue of the maximal velocity in special relativity.
We derive the dependence of mass, length, time, Doppler effect, \v Cerenkov effect and  transition radiation angle on acceleration as an analogue phenomena in special theory of relativity. The last application of our method is the Thomas precession by uniform acceleration with the possible role in the modern physics and cosmology.
The comparison of derived results with other relativistic methods is necessary. 
\end{abstract}

\vspace{15mm}

\baselineskip 13pt

\section{Introduction}

Thomas precession, named after Llewellyn Thomas, is a relativistic motion of a particle following a curvilinear orbit. Algebraically, it is a result of the non-commutativity of Lorentz transformations.
Thomas precession is a kinematic effect in the flat spacetime of special relativity.
This rotation is called Thomas rotation, Thomas and Wigner rotation or
Wigner rotation. The rotation was discovered by Thomas (Thomas, 1926) 
and derived by Wigner (Wigner, 1939). If a
sequence of non-collinear boosts returns an object to its initial velocity, then the sequence of Wigner rotations can combine to produce a net rotation called the Thomas precession (Rhodes et al., 2005). 

Thomas precession is always accompanied by dynamical effects (Malykin, 2006). 
We calculate here Thomas precession caused by accelerated  motion of the systems.
In other words we show that Thomas precession can be initiated by acceleration of a point particle.
The problem of acceleration of charged particles or systems of particles
is the permanent and the most prestige problem in the accelerator
physics. Particles can be accelerated by different ways. Usually by
the classical electromagnetic fields, or, by light pressure
of the laser fields (Baranova et al., 1994; Pardy, 1998, 2001, 2002).
The latter method is the permanent problem of the laser physics for
many years.

In the next section we determine transformations between coordinate systems
which move mutually with the same acceleration. We determine transformations
between non relativistic and relativistic uniformly accelerated systems. 

Let us remind, at first,  some ingredients of the special theory of relativity. The Lorentz transformation between two inertial coordinate
systems $S(0, x, y, z)$ and $S'(0, x', y', z')$ (where system $S'$ moves in
such a way that $x$-axes converge, while $y$ and $z$-axes run parallel and at
time $t = t' = 0$ for the origin of the systems $O$ and $O'$ it is $O
\equiv O'$) is as follows:

$$x' = \gamma(v)(x - vt), \quad y' = y,\quad z' = z',
\quad t' = \gamma(v)\left(t - \frac
{v}{c^{2}}x\right),\eqno(1)$$
where

$$\gamma(v) = \left(1 - \frac {v^{2}}{c^{2}}\right)^{-1/2}.\eqno(2)$$

The infinitesimal form of this transformation is evidently given by
differentiation of the every equation. Or,

$$dx' = \gamma(v)(dx - vdt), \quad dy' = dy,\quad dz' = dz,
\quad dt' = \gamma(v)\left(dt - \frac{v}{c^{2}}dx\right).\eqno(3)$$

It follows from eqs. (3) that  if $v_{1}$ is velocity of the inertial system 1 with regard to $S$ and $v_{2}$ is the velocity of the inertial systems 2 with regard to 1, then the relativistic sum of the two velocities is

$$v_{1} \oplus v_{2} = \frac{v_{1} + v_{2}}{1 + \frac{v_{1}v_{2}}{c^{2}}}.\eqno(4)$$

The infinitesimal form of Lorentz transformation  (3) evidently gives
the Lorentz length contraction and time dilation. Namely,
if we put $dt = 0$ in the first equation of system (3), then
the Lorentz length contraction follows
in the infinitesimal form $dx' = \gamma(v)dx$. Or, in other words,
if in the system $S'$
the infinitesimal length is $dx'$, then the relative length with
regard to the system
$S$ is $\gamma^{-1}dx'$. Similarly, from the
last equation of (3) it follows the time dilatation for $dx = 0$.
Historical view on this effect is in the Selleri article (Selleri, 1997).

\section{Uniformly accelerated frames with space-time symmetry}

Let us take two systems $S(0, x, y, z)$ and $S'(0, x', y', z')$,
where system $S'$ moves in
such a way that $x$-axes converge, while $y$ and $z$-axes run parallel and at
time $t = t' = 0$ for the beginning of the systems $O$ and $O'$ it is $O
\equiv O'$. Let us suppose that system $S'$ moves relative to some basic
system $B$ with  a positive acceleration  and system $S'$ moves relative to system $B$
with the negative acceleration. It means that both systems moves one another
with acceleration $a$ and are equivalent because
in every system it is possibly to observe the force caused by the
same acceleration. In other words no system is inertial.

Now, let us consider the formal transformation equations between two systems.
At this moment we give to this transform only formal meaning because at this
time, the physical meaning of such transformation is not known.
On the other hand,
the consequences of the transformation will be shown very interesting. The first published derivation of such transformation by the standard way was given by author (Pardy, 2003; 2004; 2005), and the same transformations were submitted some decades ago (Pardy, 1974). The old results can 
be obtained if we perform transformation 

$$t \rightarrow t^{2},\quad t' \rightarrow t'^{2},
 \quad  v \rightarrow
\frac {1}{2}a, \quad c \rightarrow \frac {1}{2}\alpha\eqno(5)$$ 
in the original Lorentz transformation (1).
We get: 

$$x' = \Gamma(a)(x - \frac {1}{2}at^{2}), \quad y' = y,\quad z' = z,
\quad t'^{2} = \Gamma(a)\left(t^{2} - \frac{2a}{\alpha^{2}}x\right)
\eqno(6)$$
with

$$\Gamma(a) = \frac {1}{\sqrt{1 - \frac {a^{2}}{\alpha^{2}}}}.\eqno(7)$$

We used  practically new denotation
of variables in order to get the transformation  (6) between accelerated systems.

The transformations (6) form the one-parametric group with the parameter $a$. The proof of this mathematical statement can be easy performed if we perform the transformation  $T_{1}$ from $S$ to $S'$, transformation  $T_{2}$  from $S'$ to $S''$ and transformation
$T_{3}$ from $S$ to $S''$. Or,

$$x' = x'(x,t; a_{1}), \quad t' = t'(x,t; a_{1}),\eqno(8)$$

$$x'' = x''(x',t'; a_{2}), \quad t'' = t''(x',t'; a_{2}).\eqno(9)$$

After insertion of transformations (8) into (9), we get 

$$x'' = x''(x,t; a_{3}), \quad t'' = t''(x,t; a_{3}),\eqno(10)$$
where parameter $a_{3}$ is equal to 

$$a_{3} = \frac {a_{1} + a_{2}}{1 + \frac {a_{1}a_{2}}{\alpha^{2}}}.
\eqno(11)$$

The inverse parameter is $-a$ and parameter for identity is $a = 0$.
It may be easy to verify that the final relation for the definition of the continuous group transformation is valid for our transformation. Namely (Eisenhart, 1943):

$$\left(T_{3}T_{2}\right)T_{1} = T_{3}\left(T_{2}T_{1}\right).\eqno(12)$$

The physical interpretation of this nonlinear transformations is the
same as in the case of the Lorentz transformation only the physical
interpretation of the invariant function $x = (1/2)\alpha t^{2}$ is
different. Namely it can be expressed by the statement: if there is a physical signal in the system $S$ with the law $x = (1/2)\alpha t^{2}$, then in the system $S'$ the law of the process is $x' = (1/2)\alpha t'^{2}$, where $\alpha$ is the constant of maximal acceleration. It is new constant and cannot be constructed from the known physical constants.

Let us remark, that it follows from history of physics, that Lorentz
transformation was taken first as physically meaningless mathematical object by Larmor, Voigt and  Lorentz  and later only Einstein decided to put the physical meaning
to this transformation and to the invariant function $x = ct$.
We hope that the derived transformation will appear as physically meaningful.

Using relations $t \leftarrow t^{2},\quad t' \leftarrow t'^{2},
 \quad  v \leftarrow \frac {1}{2}a, \quad c \leftarrow \frac {1}{2}\alpha$, the nonlinear
transformation is expressed as the Lorentz transformation
forming the one-parametric group. This proof is equivalent to the proof by the above direct calculation. The integral part of the group properties is the so called
addition theorem for acceleration.

$$a_{3} = a_{1} \oplus a_{2} = \frac {a_{1} + a_{2}}{1 + \frac {a_{1}a_{2}}{\alpha^{2}}}.
\eqno(13)$$
where $a_{1}$ is the acceleration of the $S'$ with regard to the system $S$,
$a_{2}$ is the acceleration of the system $S''$ with regard to the system
$S'$ and $a_{3}$ is the acceleration of the system $S''$  with regard to the
system $S$. The relation (13),
expresses the law of acceleration addition theorem
on the understanding that the events are marked according to the relation  (6).

If $a_{1}= a_{2} = a_{3} = .... + a_{i} = a$,  for $i$-th accelerated carts which rolls in such a way that the first cart rolls on the basic cart, the second rolls on the first cart and so on,  
then we get for the sum of $i$ accelerated carts the following formula

$$a_{sum} =   \frac{1 - \left(\frac{1 - a/\alpha }{1 + a/\alpha}\right)^{i}}
{1 + \left(\frac{1 - a/\alpha}{1 + a/\alpha}\right)^{i}},\eqno(14)$$
which is an analogue of the formula for the inertial systems (Lightman et al., 1975).
 
 In this formula as well as in the transformation equation (6) appears constant $\alpha$ which is the constant of maximal acceleration and which cannot be calculated from the theoretical considerations, or, constructed from the known physical constants (in analogy with the velocity of light). 
What is its magnitude  can be established only by  experiments. The notion
maximal acceleration was introduced some  decades ago by author (Pardy, 1974).
 Caianiello (1981) introduced it as some consequence
of quantum mechanics and Landau theory of fluctuations. Revisiting view
on the maximal acceleration was given by Papini (2003).
At present time it was argued by Lambiase et al. (1999) that
maximal acceleration determines the upper limit of the Higgs boson and that it gives also the relation which links the mass of $W$-boson with the mass of the Higgs boson.
The LHC and HERA experiments presented  different  answer to this problem.

\section{Dependence of mass, length, time, the  Doppler effect, the \v Cerenkov effect and the transition radiation angle on acceleration}

If the maximal acceleration is the physical reality, then it should have the
similar consequences in a dynamics as the maximal velocity of motion has
consequences in the dependence of mass on velocity.
We can suppose in analogy with the special relativity
that mass depends on the acceleration for small
velocities,
in the similar way as it depends on velocity in case of uniform motion.
Of course such assumption must be experimentally verified and in no case
it follows from special theory of relativity, or, general theory of
relativity (Fok, 1961). So, we postulate
ad hoc, in analogy with special theory of relativity ( $c \rightarrow \alpha/2,  v \rightarrow 
a$):

$$m(a) = \frac {m_{0}}{\sqrt{1 - \frac {a^{2}}{\alpha^{2}}}};
\quad v \ll c, \quad a = \frac{dv}{dt}.\eqno(15)$$

Let us derive as an example the law of
motion when the constant force $F$  acts on the body with the rest
mass $m_{0}$. Then, the Newton law reads (Landau et al., 1997):

$$F = \frac{dp}{dt} =
m_{0}\frac{d}{dt}\frac {v}{\sqrt{1 - \frac {a^{2}}{\alpha^{2}}}}.
\eqno(16)$$

The first integral of this equation can be written in the form:

$$\frac{Ft}{m_{0}} = \frac {v}{\sqrt{1 - \frac {a^{2}}{\alpha^{2}}}};
\quad a = \frac{dv}{dt}, \quad F = const.. \eqno(17)$$

Let us introduce quantities

$$v = y,\quad a = y', \quad A(t) =
\frac{F^{2}t^{2}}{m_{0}^{2}\alpha^{2}}.\eqno(18)$$

Then, the quadratic form of the equation (17) can be written as
the following differential equation:

$$A(t)y'^{2} + y^{2} - A(t)\alpha^{2} = 0, \eqno(19)$$
which is nonlinear differential equation of the first order.
The solution of it is of the form $y = Dt$, where $D$ is some constant,
which can be easily determined. Then, we have the solution in the form:

$$y = v = Dt = \frac {t}{\sqrt{\frac {m_{0}^{2}}{F^{2}} + \frac {1}{\alpha^{2}}}}.
\eqno(20)$$

For F $\rightarrow \infty$, we get $v = \alpha t$. This relation can
play substantial role at the beginning of the big-bang, where the accelerating
forces can be considered as infinite, however the law of acceleration
has finite nonsingular form.

At this moment it is not clear if the dependence of the mass on acceleration
can be related to the energy dependence on acceleration similarly to the Einstein relation uniting energy, mass and velocity (Okun, 2001, 2002; Sachs, 1973 ).

The infinitesimal form of author transformation  (6) evidently gives
the  length contraction and time dilation. Namely,
if we put $dt = 0$ in the first equation of system (6), then
the  length contraction follows
in the infinitesimal form $dx' = \Gamma(a)dx$. Or, in other words,
if in the system $S'$
the infinitesimal length is $dx'$, then the relative length with
regard to the system
$S$ is $\Gamma^{-1}dx'$. Similarly, from the
last equation of (6) it follows the time dilatation for $dx = 0$.

The relativistic Doppler effect is the change in frequency (and wavelength) of light, caused by the relative motion of the
source and the observer (as in the classical Doppler effect), when taking into account effects described by the special theory of
relativity.

The relativistic Doppler effect is different from the non-relativistic Doppler effect as the equations include the time dilation effect of special relativity and do not involve the medium of propagation as a reference point (Rohlf, 1994).

The Doppler shift caused by acceleration can be also derived immediately from the original relativistic equations for the Doppler shift. We only make the transformation $v \rightarrow a/2, c \rightarrow \alpha/2$ to get 

$$\frac{\lambda'}{\lambda} = \sqrt{\frac{1 - a/\alpha}{1 + a/\alpha}},\eqno(21)$$
when the photons of the wave length $\lambda$ are measured toward photon source, and 

$$\frac{\lambda'}{\lambda} = \sqrt{\frac{1 + a/\alpha}{1 - a/\alpha}},\eqno(22)$$
when the photons of the wave length $\lambda$ are measured in the frame that is moving away from the photon source. Different approach used Friedman et al. (2010).

Concerning the \v Cerenkov radiation, it is based on the fact that the speed of light in the medium with the index of refraction $n$ is $c/n$. A charged particle moving in such medium can have the speed greater than it is the speed of light in this medium. When a charged particle moves faster than the speed of light in this medium, a portion of the electromagnetic radiation emitted by excited atom along the path of the particle is coherent. The coherent radiation is emitted at a fixed angle with respect to the particle trajectory. This radiation was observed by \v Cerenkov in 1935. The characteristic angle was derived by Tamm and Frank in the form (Rohlf, 1994)

$$\cos\theta = \frac{c}{vn}.\eqno(23)$$ 

The \v Cerenkov angle  caused by acceleration can be also derived immediately from the original  Frank-Tamm equations for 
this effect. We only make the transformation $v \rightarrow a/2, c \rightarrow \alpha/2$ to get 

$$\cos\theta = \frac{\alpha}{an}.\eqno(24)$$ 

In case of the Ginzburg transition radiation, the radiation in concentrated in the angle 

$$1/\gamma = \frac{1}{\sqrt{1 - \frac{v^{2}}{c^{2}}}}.\eqno(25)$$ 

The transition radiation angle  caused by acceleration can be also derived immediately from the original  
Ginzburg formula  for this effect. We only make the transformation $v \rightarrow a/2, c \rightarrow \alpha/2$ to get 

$$1/\Gamma = \frac{1}{\sqrt{1 - \frac{a^{2}}{\alpha^{2}}}}.\eqno(26)$$ 

Let us remark that all formulas derived in this section  involving the uniform acceleration $a$ can be used for the uniform equivalent gravity according to  the principle of equivalence.
 
\section{Thomas precession in uniformly accelerated system}

With regard to the fact that  new results in uniformly accelerated systems can be obtained from  the old relativistic results having the form of the mathematical objects involving function $f(v/c)$ we use this algorithm to derive the Thomas angle from original relativistic angle with transformation  $t \leftarrow t^{2},\quad t' \leftarrow t'^{2},  \quad  v \leftarrow \frac {1}{2}a, \quad c \leftarrow \frac {1}{2}\alpha$.

However, let us at first remind the relativistic derivation of the Thomas angle.
 So, let us consider the inverse transformations of (6) with  $T_{1}$ from $S$ to $S'$ with velocity ${\bf v} || x$,
transformation  $T_{2}$  from $S'$ to $S''$ with velocity ${\bf u} || y$ and transformation
$T_{3}$ from $S$ to $S^{+}$ with velocity ${\bf v} \oplus {\bf u} $ where  mathematical symbol $\oplus$ is the expression for the relativistic addition of the velocities  ${\bf v},{\bf u} $. Then $S = S^{+}$, if $S^{+}$ is turned in the xy plane with angle $\varphi$, which is given by the formula (Tomonaga, 1997):

$$\varphi = 
\arctan \frac{uv(\sqrt{1-\frac{u^{2}}{c^{2}}} \sqrt{1-\frac{v^{2}}{c^{2}}} -1)}
{u^{2}\sqrt{1-\frac{v^{2}}{c^{2}}} + v^{2}\sqrt{1-\frac{u^{2}}{c^{2}}}}.
\eqno(27)$$

To see it let us perform the transformations properly. Let be 
$S\rightarrow S', \quad {\bf v} = (v,0,0)$, transformation. Or, 

$$x = \gamma_{v}(x' + vt'), \quad  y = y',\quad 
t = \gamma_{v}(t' + \frac{v}{c^{2}}x');\quad \gamma_{v} =\left( 1 - \frac{v^{2}}{c^{2}}\right)^{-1/2}.\eqno(28)$$

Then, let be $S'\rightarrow S'', \quad {\bf u} = (0,u, 0)$. Or, 

$$x' = x'',\quad y' = \gamma_{u}(y'' + ut''),  \quad t' = \gamma_{u}(t'' + \frac{u}{c^{2}}y'');\quad \gamma_{u} =\left( 1 - \frac{u^{2}}{c^{2}}\right)^{-1/2}.
\eqno(29)$$

The transformation from $S$ to $S''$ is $S \rightarrow S''$. Or, 

$$x = \gamma_{v}x''  + \gamma_{v}\gamma_{u}\frac{vu}{c^{2}}y'' + \gamma_{v}\gamma_{u}vt'',\quad 
y = \gamma_{u}y'' + \gamma_{u}u t'', \eqno(30) $$

$$t = \gamma_{v}\frac{v}{c^{2}}x'' +  \gamma_{v}\gamma_{u}\frac{u}{c^{2}}y'' + \gamma_{v}\gamma_{u}t''.\eqno(31)$$

Now, let us perform transformation from $S$ to $S^{+}$, where $S^{+}$ moves with
regard to $S$ with velocity, which is the relativistic sum of $\bf v$ and $\bf u$, which is the  velocity  ${\bf v} \oplus {\bf u}$. Or, using formula (Batygin et al., 1970)

$$ {\bf k} = {\bf v} \oplus {\bf u}  = \frac{{\bf v} + {\bf u} + 
\left(\gamma_{v} -1\right)\frac{{\bf v}}{v^{2}}
\left[{\bf v}\cdot{\bf u} + v^{2}\right]}
{\gamma_{v}\left(1 + \frac{{\bf v}\cdot{\bf u}}{c^{2}}\right)},\eqno(32)$$
we get

$${\bf k} = {\bf v} \oplus {\bf u} = \left(v, \frac{u}{\gamma_{v}}, 0\right);\quad 
\gamma_{k} =\left( 1 - \frac{k^{2}}{c^{2}}\right)^{-1/2}.\eqno(33)$$
 
Then we have for radius vector ${\bf r}$ and time $t$ we have transformations (Batygin et al., 1970):

$$\widetilde{{\bf r}}  = {\bf r}^{+} + {\bf k}t^{+} + \left(\gamma_{k} -1\right)\frac{{\bf k}}{k^{2}}
\left[{\bf k}{\bf r}^{+} + k^{2}t^{+}\right]\eqno(34)$$
and 

$$\widetilde{t} = \gamma_{k}\left(t^{+} + \frac{{\bf k}{\bf r}^{+}}{c^{2}}\right).\eqno(35)$$

The t-transformation (35) can be expressed in variables $t^{+},x^{+},y^{+}$ as follows:

$$\widetilde{t} = \gamma_{k}t^{+} + \gamma_{k}\frac{ v}{c^{2}}x^{+} + \frac{\gamma_{u}}{\gamma_{v}}\frac{u}{c^{2}}y^{+}.\eqno(36)$$

The last equation can be compared with the time transformation from $S$ to $S''$, which is 

$$t = \gamma_{v}\gamma_{u}t'' +  
\gamma_{v}\frac{v}{c^{2}}x'' + \gamma_{v}\gamma_{u}\frac{u}{c^{2}}y''.\eqno(37)$$

Using $\gamma_{k} = \gamma_{v}\gamma_{u}$, we get two transformation of time following from eqs. (31) and (36):

$$S \rightarrow S'':  t = \gamma_{k}t'' +  
\gamma_{v}\frac{v}{c^{2}}x'' + \gamma_{k}\frac{u}{c^{2}}y'',\eqno(38)$$

$$S \rightarrow S^{+}:\widetilde{t} = \gamma_{k}t^{+} + \gamma_{k}\frac{ v}{c^{2}}x^{+} + 
\frac{\gamma_{k}}{\gamma_{v}}\frac{u}{c^{2}}y^{+}.\eqno(39)$$

Now, let us perform rotation 

$$x'' = x^{+}\cos\varphi + y^{+}\sin\varphi, 
 \quad y'' = - x^{+}\sin\varphi + y^{+}\cos\varphi. \eqno(40)$$ 

Then equation (39) is identical with eq. (40), if the angle $\varphi$ is determined by equation

$$\varphi = \arctan\frac{\left(1 - \gamma_{v}\gamma_{u}\right)vu}
{\gamma_{v}v^{2} +\gamma_{u}u^{2}}. \eqno(41)$$ 

The angle of rotation (41) is so called the Thomas angle of so called Thomas precession.  
With regard to the derived transformation of quantities ${\bf u},{\bf v}, c$  to  the uniformly accelerated system, or, ${\bf v} \rightarrow {\bf a}/2 $, ${\bf u
} \rightarrow {\bf w}/2 $ $c \rightarrow \alpha/2 $, we get immediately from the last formula (41) the Thomas precession angle:

$$\varphi = 
\arctan \frac{aw(\sqrt{1-\frac{a^{2}}{\alpha^{2}}} \sqrt{1-\frac{w^{2}}{\alpha^{2}}} -1)}{a^{2}\sqrt{1-\frac{w^{2}}{\alpha^{2}}} + w^{2}\sqrt{1-\frac{a^{2}}{\alpha^{2}}}},
\eqno(42)$$
which has the physical meaning of the Thomas precession  caused by uniform acceleration. 
The last formula with uniform acceleration $a$  and $w$ can be used for the uniform equivalent gravity according to  the principle of equivalence.
It is not excluded that this formula will play the crucial role in modern physics with application for LHC in CERN. 
 
\section{Discussion}

The maximal acceleration constant which was introduced here is kinematic one and it differs 
from the Caianiello (1981) definition following from quantum mechanics.
Our constant cannot be determined by the system of other physical constants. It is an analogue of the numeric velocity of light which cannot be composed from other physical constants, or,
the Heisenberg fundamental length in particle physics.
The nonlinear transformations (13) changes the Minkowski metric

$$ds^{2} = c^{2}dt^{2} - dx^{2} - dy^{2} - dz^{2}\eqno(43)$$ 
to the new metric with the Riemann form. Namely ($c \rightarrow \alpha/2, dt \rightarrow 2tdt$):

$$ds^{2} = \alpha^{2}t^{2}dt^{2} - dx^{2} - dy^{2} - dz^{2}\eqno(44)$$ 
and it can be investigated by the methods of differential geometry. 
So, equations (13) and (51)
can form the preamble to investigation of accelerated systems. 

If some experiment will confirm the existence of kinematical maximal
acceleration $\alpha$, then it will have
certainly crucial consequences for Einstein theory of gravity
because this theory does not involve this factor.
Also the cosmological theories constructed on the basis of the
original Einstein equations will require modifications. The so called Hubble constant 
will be changed  and the scenario of the accelerating universe modified.

Also the standard model of  particle physics and supersymmetry  theory
will require generalization because they  does not involve the maximal acceleration constant.
It is not excluded that also the theory of parity nonconservation will be modified by the maximal acceleration constant. In such a way the  particle laboratories have perspective applications involving the physics with maximal acceleration. Many new results can be obtained from  the old relativistic results having the form of the mathematical objects involving function $f(v/c)$.
The derived  formulas with  uniform acceleration $a$  and $w$ can be applied and verified in case 
of the uniform equivalent gravity according to the principle of equivalence.

The prestige problem in the modern theoretical physics -
the theory of the Unruh effect, or, the existence of thermal radiation detected by accelerated  observer -  is in the development (Fedotov et al., 2002) and the
serious statement, or comment to the relation of this effect to the maximal acceleration must be elaborated. The analogical statement is valid for the Hawking effect in the theory of black holes.

It is not excluded that the maximal acceleration constant 
will be discovered by ILC. The unique feature of the International Linear Collider (ILC) is the fact that its CM energy can be increased gradually simply by extending the main linac.

Let us remark in conclusion that it is possible to extend and  modify quantum field theory by maximal acceleration. It is not excluded that the kinematic maximal acceleration constant  will enable to reformulate the theory of renormalization. 

\vspace{10mm}

{\bf REFERENCES}

\vspace{7mm}

\noindent
Baranova, N. B.  and  Zel'dovich, B. Ya. (1994).  Acceleration of\\
charged particles by laser beams,  JETP {\bf 78}, (3),  249-258.\\[2mm] 
Batygin, V. V. and Toptygin, I. N. {\it Collections of problems in electrodynamics},
(Nauka, Moscow, 1970). (in Russian).\\ [2mm]
Caianiello, E. R. (1981).  Is There maximal acceleration? 
Lett. Nuovo Cimento,  {\bf 32}, 65 ; ibid. (1992).
Revista del Nuovo Cimento, {\bf 15}, No. 4. \\[2mm]
Eisenhart, L. P. {\it Continuous groups of transformations},  (Princeton, 1943).  \\[2mm]
Fedotov, A. M.,  Narozhny, N. B. ,  Mur, V. D. and Belinski,  V. A. (2002).  
An example of a uniformly accelerated particle detector with non-Unruh response, arXiv:
hep-th/0208061. \\[2mm]
Fok, V. A. {\it The Theory of space, time and gravity},
second edition, (GIFML, Moscow, 1961). (in Russian). \\[2mm]
Friedman, Y. (2010). The maximal acceleration, extended relativistic
dynamics and Doppler type shift for an accelerated source, arXiv:0910.5629v3
[physics.class-ph] 19 Jul 2010. \\[2mm]
Lambiase, G.,  Papini, G. and Scarpetta, G. (1998). Maximal acceleration limits on the mass of the Higgs boson, arXiv:hep-ph/9808460; ibid. (1999). Nuovo Cimento B, 
{\bf 114}, 189-197. \\[2mm]
Lightman, A. P., Press W. H., Price, R. H. and Teukolsky S. A. {\it Problem book in relativity and gravitation},  (Princeton University Press, 1975 ). \\ [2mm]
Malykin, G. B. (2006). Thomas precession: correct and incorrect solutions, 
Phys. Usp. {\bf 49}, 83 (2006).\\ [2mm]
Okun, L. B. (2001). Photons, clocks, gravity and concept of mass, arXiv:physics/0111134v1, 
ibid. (2002). Nucl. Phys. Proc. Suppl. {\bf 110},  151-155. \\[2mm]
Papini, G. (2003). Revisiting Caianiello's maximal acceleration, e-print quant-ph/0301142. \\[2mm]
Pardy, M. (1974). The group of transformations for accelerated systems, ( Department of Theoretical Physics, Jan Evangelista Purkyn\v e University, Now, Masaryk University), Brno (unpublished). \\[2mm] 
Pardy, M. (1998).  The quantum field theory of laser acceleration,
Phys. Lett. A {\bf 243},  223-228. \\[2mm]
Pardy, M. (2001). The quantum electrodynamics of laser acceleration,
Radiation Physics and Chemistry {\bf 61},  391-394.\\[2mm]
Pardy, M. (2002). Electron in the ultrashort laser pulse, arXiv:hep-ph/02372; ibid.
International Journal of Theor. Physics, {\bf 42}, No.1, 99. \\[2mm]
Pardy, M. (2003).  The space-time transformation and the maximal acceleration, arXiv:grqc/
0302007. \\[2mm]
Pardy, M. (2004). The space-time transformation and the maximal acceleration,
Spacetime \& Substance Journal, {\bf 1}(21), (2004), pp. 17-22. \\[2mm]
Pardy, M. (2005). Creativity leading to discoveries in particle physics and relativity,
arXiv:physics/0509184 \\[2mm]
Rohlf, J. W.  {\it Moderm physics from $\alpha$ to $Z^{0}$}
(John Wiley \& Sons, Inc. New York, 1994). \\[2mm]
Rhodes, J. A. and  Semon, M. D. (2005).  Relativistic velocity space, Wigner rotation and Thomas precession, arXiv:gr-qc/0501070v1 24 Jan 2005. \\[2mm]
Sachs, M. (1973). On the Meaning of $E = mc^{2}$, International Journal of
Theoretical Physics, {\bf 8}, No. 5, 377 (1973). \\[2mm]
Selleri, F. (1997). The relativity principle and nature of time, 
Foundations of Physics, {\bf 27}, No. 11, 1527. \\[2mm]
Thomas, L. H. (1926). Motion of the spinning electron", Nature {\bf 117}, 514. \\[2mm]
Tomonaga, Sin-intiro. {\it The story of spin}, (University of Chicago Press, 1997). \\[2mm]
Wigner, E. P. (1939). On unitary representations of the inhomogeneous Lorentz group,
Ann. Math. {\bf 40}, 149--204.
\end{document}